\documentstyle[12pt]{article}

\newcommand{\vr}{{\bf r}}
\newcommand{\dlt}{\delta}
\newcommand{\om}{\omega}

\newcommand{\vp}{\varphi}
\newcommand{\be}{\begin{equation}}
\newcommand{\ee}{\end{equation}}
\newcommand{\prt}{\partial}
\newcommand{\ra}{\rightarrow}

\begin{document}

\begin{center}
{\Large{\bf Spectrum of Coherent Modes for Trapped Bose Gas} \\ [5mm]
V.I. Yukalov$^{1,2}$, E.P. Yukalova$^{1,3}$, and V.S. Bagnato$^{1}$}\\ [2mm]

{\it 
$^1$Instituto de Fisica de S\~ao Carlos, Universidade de S\~ao Paulo \\
Caixa Postal 369, S\~ao Carlos, S\~ao Paulo 13560-970, Brazil \\ [2mm]

$^2$Bogolubov Laboratory of Theoretical Physics \\
Joint Institute for Nuclear Research, Dubna 141980, Russia \\ [2mm]

$^{3}$Department of Computational Physics \\
Laboratory of Informational Technologies \\
Joint Institute for Nuclear Research, Dubna 141980, Russia \\ [5mm]
E-mail: yukalov@if.sc.usp.br}
\end{center}

\vskip 3cm

\begin{abstract}

The spectrum of coherent modes for dilute atomic Bose gas, confined 
in a cylindrical trap, is calculated by applying optimized perturbation
theory and the technique of self-similar root approximants. The latter
technique makes it possible to derive accurate analytical formulas.
The obtained expressions are valid for arbitrary energy levels of all 
excited nonlinear coherent modes and for different traps, spherical,
sigar-shape, and disk-shape.

\end{abstract}

\newpage

\section{Introduction}

At low temperatures, Bose-Einstein condensate of trapped atomic gases 
is well described by the Gross-Pitaevskii equation (see reviews [1--3]). 
In equilibrium, the condensate state corresponds to the ground-state 
solution of this equation. The mathematical structure of the 
Gross-Pitaevskii equation is identical to the nonlinear Schr\"odinger
equation, which in the presence of a confining potential,
should possess a discrete set of stationary states. Such states can be
interpreted [4] as {\it nonlinear coherent modes} of trapped atoms [4]
corresponding to {\it nonground-state Bose condensates}. The excitation
of these modes can be accomplished by means of resonant modulation of an
external potential [4--6]. Such modes have also been studied in Refs.
[7--9] and a dipole mode has been observed experimentally [10].

One should not confuse the nonlinear coherent modes, that are stationary
solutions to the {\it nonlinear} Gross-Pitaevskii equation, with 
elementary collective excitations, which are solutions of the {\it linear}
Bogolubov-De Gennes equations. The nonlinear coherent modes are sometimes
called {\it topological modes} to stress that their spatial behaviour
qualitatively differs them from each other. Each nonlinear mode generates
its own collective excitations as small deviations around the given mode.
Usually, one considers collective excitations above the ground-state
mode. These excitations correspond to small density fluctuations and
are in a reasonable agreement with the Bogolubov spectrum for both 
superfluid helium [11] as well as for trapped Bose-Einstein condensate
[12]. In the first case, it is the phonon-roton spectrum, which is 
a unified branch [13,14]. In the case of dilute trapped atoms, it is a 
phonon-single-particle branch. Collective excitations around nonground 
state coherent modes have not yet been considered.

Before considering perturbations of nonlinear coherent modes, it is 
necessary to have an accurate description of their own properties. In
particular, one needs to better understand the features of the spectrum
of the nonlinear modes themselves. The study of this spectrum was
initiated in Ref. [4] and, for a weak interaction, was also considered
in Ref. [9]. In the present paper, we aim at giving a general description 
of the nonlinear-mode spectrum for trapped atoms in a cylindrical trap.
We shall describe a method for calculating arbitrary energy levels of
this spectrum for any magnitude of the interaction parameter. We shall
also construct accurate analytical expressions for the nonlinear-mode
spectrum. Having in hands such analytical expressions is convenient for
studying the spectrum dependence on quantum numbers and system parameters
as well as for an easier comparison with experiments.

\section{Equation for Coherent Modes}

The interaction of atoms in dilute gases is characterized by the Fermi
contact potential
$$
\Phi(\vr) = A\dlt(\vr) \; , \qquad A\equiv 4\pi\hbar^2\;
\frac{a_s}{m_0} \; ,
$$
in which $a_s$ is the $s$-wave scattering length and $m_0$, the atomic 
mass. The trapping potential is usually taken in the form
$$
U(\vr) = \frac{m_0}{2}\; \left ( \om_x^2 r_x^2 + \om_y^2 r_y^2 +
\om_z^2 r_z^2\right ) \; .
$$
The stationary coherent field $\vp(\vr)$, normalized to unity, is 
described [3] by the equation
\be
\label{1}
\hat H[\vp(\vr)]\vp(\vr) = E\vp(\vr) \; ,
\ee
which is often called the Gross-Pitaevskii equation, with the nonlinear
Hamiltonian
\be
\label{2}
\hat H[\vp(\vr)] = -\; \frac{\hbar^2{\bf\nabla}^2}{2m_0} +
U(\vr) + NA |\vp(\vr)|^2 \; .
\ee
Because of the confining potential $U(\vr)$, the eigenproblem (1) 
possesses a discrete spectrum, with the related eigenfunctions being 
{\it nonlinear coherent modes} [4].

In what follows, we shall consider a harmonic potential of
cylindrical symmetry, with a radial frequency
\be
\label{3}
\om_\perp \equiv \om_x= \om_y
\ee
and the anisotropy parameter
\be
\label{4}
\nu \equiv \frac{\om_z}{\om_\perp} \; .
\ee
It is convenient to work with dimensionless quantities, measuring the
space variables
\be
\label{5}
r \equiv \frac{\sqrt{r_x^2+ r_y^2}}{l_\perp} \; , \qquad
z\equiv \frac{r_z}{l_\perp} 
\ee
in units of the transverse oscillator length
\be
\label{6}
l_\perp \equiv \sqrt{\frac{\hbar}{m_0\om_\perp}} \; .
\ee
The coupling parameter
\be
\label{7}
g\equiv 4\pi\frac{a_s}{l_\perp} \; N
\ee
is a dimensionless quantity characterizing atomic interactions. Defining
the dimensionless Hamiltonian and wave function, respectively,
\be
\label{8}
\hat H \equiv \frac{\hat H[\vp(\vr)]}{\hbar\om_\perp} \; , \qquad
\psi(r,\vp,z) \equiv l_\perp^{3/2}\vp(\vr) \; ,
\ee
in the cylindrical coordinates, we have
\be
\label{9}
\hat H = -\; \frac{1}{2}\; {\bf\nabla}^2 + \frac{1}{2}\left ( r^2 +
\nu^2 z^2 \right ) + g |\psi|^2 \; ,
\ee
where
$$
{\bf\nabla}^2 = \frac{\prt^2}{\prt r^2} + \frac{1}{r}\; 
\frac{\prt}{\prt r} + \frac{1}{r^2}\; \frac{\prt^2}{\prt\vp^2} +
\frac{\prt^2}{\prt z^2} \; .
$$
Then the eigenproblem (1) transforms to the equation
\be
\label{10}
\hat H\psi_n = E_n\psi_n \; ,
\ee
defining dimensionless coherent modes $\psi_n$ and their spectrum 
$\{ E_n\}$; $n$ being a multi-index labelling the modes. The latter 
are assumed to be normalized as
$$
\int_0^\infty r\; dr \; \int_0^{2\pi} d\vp\;
\int_{-\infty}^{+\infty} dz\; |\psi_n(r,\vp,z)|^2 =  1 \; .
$$
Our aim is to find the spectrum $\{ E_n\}$ for all energy levels $E_n$ 
of any quantum index $n$ and for arbitrary coupling and anisotropy 
parameters $g$ and $\nu$, respectively.

\section{Optimized Perturbation Theory}

To calculate the spectrum of the eigenproblem (10), we shall employ the 
optimized perturbation theory [15--19], which has been successfully
applied to a number of models in quantum mechanics, statistical physics,
and quantum field theory [15--26]. Numerous references on various 
applications can be found in surveys [27--29]. It is important to stress
that the optimized perturbation theory was shown to provide accurate
results for all energy levels and large coupling parameters.

For the case considered, the procedure can be as follows. We start with 
the initial Hamiltonian of a harmonic oscillator
\be
\label{11}
\hat H_0 = - \; \frac{1}{2}\; {\bf\nabla}^2 + \frac{1}{2}\left (
u^2 r^2 + v^2 z^2\right ) \; ,
\ee
having the oscillator strengths as two trial parameters, $u$ and $v$. 
This Hamiltonian possesses the eigenvalues
\be
\label{12}
E_{nmj}^{(0)} = ( 2n +|m|+1) u +\left ( j +\frac{1}{2}\right ) v \; ,
\ee
with the radial quantum number $n=0,1,2,\ldots$, asimuthal number 
$m=0,\pm 1,\pm 2,\ldots$, and the axial quantum number $j=0,1,2,\ldots$.
The corresponding wave functions are
$$
\psi_{nmj}^{(0)}(r,\vp,z) = \left [ \frac{2n!u^{|m|+1}}{(n+|m|)!}
\right ]^{1/2} r^{|m|}\exp\left ( -\; \frac{u}{2}\; r^2\right )\times
$$
\be
\label{13}
\times
L_n^{|m|}\left ( ur^2\right ) \frac{e^{im\vp}}{\sqrt{2\pi}} \left (
\frac{v}{\pi}\right )^{1/4} \frac{1}{\sqrt{2^j j!}}\; 
\exp\left ( -\; \frac{v}{2}\; z^2\right ) H_j(\sqrt{v}\; z) \; ,
\ee
where $L_n^m(\cdot)$ is a Laguerre polynomial and $H_j(\cdot)$ is a
Hermite polynomial.

Using a variant of the standard perturbation theory, such as the
Rayleigh-Schr\"odinger theory, we may find a sequence $\{ E_k(g,u,v)\}$ 
of the energies $E_k(g,u,v)$ for the approximation orders 
$k=0,1,2,\ldots$. Here, for the simplicity of notation, the dependence 
of the energy levels on the quantum numbers is not written down 
explicitly, but it is assumed. For instance, $E_0(g,u,v)=E_{nmj}^{(0)}$.
Then the trial parameters are to be transformed to control functions
[15--19] $u_k(g)$ and $v_k(g)$ such that to render the sequence 
$\{ e_k(g)\}$ of the {\it optimized terms}
\be
\label{14}
e_k(g) \equiv E_k(g,u_k(g),v_k(g))
\ee
convergent. The control functions can be found from optimization
conditions. For example, we may employ the fixed-point condition
\be
\label{15}
\left ( \dlt u\; \frac{\prt}{\prt u} + \dlt v\; \frac{\prt}{\prt v}
\right ) E_k(g,u,v) = 0 \; ,
\ee
whose solutions are $u=u_k(g)$ and $v=v_k(g)$. The {\it optimized 
approximants} (14) are valid for arbitrary quantum numbers and for the
whole range of the coupling parameter $g$. The same procedure can be 
used for anharmonic traps, described by anharmonic confining potentials 
of different powers [29], integer or noninteger. Note that the usage of 
anharmonic traps may be important for some experiments [30] dealing with 
elementary collective excitations.

In the first order, we have the energy
\be
\label{16}
E_1(g,u,v) = \frac{p}{2}\left ( u + \frac{1}{u}\right ) +
\frac{q}{4}\left ( v +\frac{\nu^2}{v}\right ) +
u\sqrt{v}\; g I_{nmj} \; ,
\ee
in which the notation
\be
\label{17}
p\equiv 2n +|m| + 1 \; , \qquad q\equiv 2j + 1
\ee
for the combinations of quantum numbers is introduced and where
$$
I_{nmj} \equiv \frac{1}{u\sqrt{v}} \int |\psi_{nmj}^{(0)}(r,\vp,z)|^4\;
rdr\; d\vp\; dz =
$$
$$
= \frac{2}{\pi^2}\left [ \frac{n!}{(n+|m|)!\; 2^j j!}
\right ]^2 \int_0^\infty x^{2|m|} e^{-2x}\left [ L_n^{|m|}(x)\right ]^4 
dx\; \int_0^\infty e^{-2t^2} H_j^4(t)\; dt \; .
$$
The fixed-point condition (15) gives the equations
\be
\label{18}
p\left ( 1 -\; \frac{1}{u^2}\right ) + \frac{s}{p\nu}\sqrt{\frac{v}{q}}
= 0 \; , \qquad q \left ( 1 -\; \frac{\nu^2}{v^2}\right ) + 
\frac{us}{p\nu\sqrt{vq}} = 0 
\ee
for the control functions $u=u_1(g)$ and $v=v_1(g)$, where, for 
convenience, the notation
\be
\label{19}
s\equiv 2p\sqrt{q}\; g\; \nu I_{nmj}
\ee
is used. Substituting the solutions for the control functions, given by
Eqs. (18), in the form (16) results in the optimized approximant $e_1(g)$,
according to definition (14). This procedure can be continued to a desired
approximation order. Here we limit ourselves by the first-order 
approximation
\be
\label{20}
E\equiv e_1(g) = E_1(g,u_1(g),v_1(g)) \; ,
\ee
in which the control functions $u_1(g)$ and $v_1(g)$ are the solutions 
of Eqs. (18). The latter equations, for a given set $\{ n,m,j\}$ of 
quantum numbers and a coupling parameter $g$, require a numerical 
solution. Hence, the spectrum (20) can be calculated only numerically. 
However, it is always desirable to possess an approximate analytical 
expression that could be easy to treat with respect to varying system 
parameters, such as quantum numbers and the coupling parameter.

\section{Self-Similar Root Approximants}

In order to derive analytical formulas for the spectrum of the 
nonlinear coherent modes, we may employ the technique of self-similar 
crossover approximants [31,32]. For this purpose, we need to know the 
asymptotic expansions of the spectrum in the limits of the weak and 
strong coupling. These expansions for the spectrum (20) can be derived 
from Eqs. (16) and (18). It is convenient to use the variable (19) 
that is proportional to the coupling parameter; so that if $g\ra 0$, 
then $s\ra 0$, and when $g\ra\infty$, then $s\ra\infty$. In the 
weak-coupling limit, we find the expansion
\be
\label{21}
E\simeq a_0 + a_1 s + a_2 s^2 + a_3 s^3 \; ,
\ee
as $s\ra 0$, where
$$
a_0 = p + \frac{q\nu}{2} \; , \qquad a_1 = \frac{1}{2p(q\nu)^{1/2}} \; ,
\qquad a_2 = -\; \frac{p+2q\nu}{16 p^3(q\nu)^2}\; , \qquad 
a_3 = \frac{(p+2q\nu)^2}{64p^5(q\nu)^{7/2}} \; .
$$
In the strong-coupling limit, we obtain
\be
\label{22}
E\simeq b_0 s^{2/5} + b_1 s^{-2/5} + b_2 s^{-6/5} + b_3 s^{-2} +
b_4 s^{-14/5} + b_5 s^{-18/5} \; ,
\ee
as $s\ra\infty$, where
$$
4b_0 = 5\; , \qquad 4b_1 = 2p^2 +(q\nu)^2 \; , \qquad
20b_2 = -3p^4 + 2p^2(q\nu)^2 - 2(q\nu)^4 \; ,
$$
$$
20b_3 = 2p^6 -p^4(q\nu)^2 -2p^2(q\nu)^4 + 2(q\nu)^6 \; , 
$$
$$
500b_4 = -44p^8+22p^6(q\nu)^2+2p^4(q\nu)^4+78p^2(q\nu)^6-69(q\nu)^8 \; ,
$$
$$
12500b_5 = 1122p^{10} -595 p^8(q\nu)^2 - 70p^6(q\nu)^4 + 440 p^4 (q\nu)^6
- 3640p^2(q\nu)^8 + 2821 (q\nu)^{10} \; .
$$

The asymptotic expansions (21) and (22), valid in the weak-coupling 
limit $s\ra 0$ and, respectively, in the strong-coupling limit 
$s\ra\infty$, can be sewed by using the technique of self-similar 
crossover approximants [31,32]. This technique makes it possible to 
construct interpolative formulas that provide correct asymptotic 
expansions, at the same time giving good accuracy in the whole interval 
of the variable $s$. The detailed description of the method can be 
found in Refs. [31,32]. Taking into account only one term in the 
strong-coupling expansion (22), we get the first-order self-similar 
root approximant
\be
\label{23}
E_1^* = a_0( 1 + As)^{2/5} \; ,
\ee
in which
$$
a_0 = p + \frac{q\nu}{2}\; , \qquad Aa_0^{5/2} = 1.746928 \; .
$$
Retaining two terms in Eq. (22) yields the second-order approximant
\be
\label{24}
E_2^* = a_0\left [ \left ( 1+ A_1 s\right )^{6/5} + A_2 s^2
\right ]^{1/5} \; ,
\ee
where $a_0$ is the same as in Eq. (23) and
$$
A_1 a_0^{25/6} = 2.533913 \left [ 2p^2 +(q\nu)^2\right ]^{5/6} \; ,
\qquad A_2 a_0^5 =  3.051758 \; .
$$
Similarly, in the third order we find
\be
\label{25}
E_3^* = a_0\left\{ \left [ \left ( 1 + B_1 s\right )^{6/5} +
B_2 s^2\right ]^{11/10} + B_3 s^3\right \}^{2/15} \; ,
\ee
where $a_0$ is again as earlier and
$$
B_1 a_0^{125/22} \left [ 2p^2 +(q\nu)^2\right ]^{5/66} =
1.405455\left [ 8p^4 + 12p^2(q\nu)^2 +(q\nu)^4\right ]^{5/6} \; ,
$$
$$
B_2 a_0^{75/11} = 6.619620\left [ 2p^2 +(q\nu)^2\right ]^{10/11} \; ,
\qquad B_3 a_0^{15/2} = 5.331202 \; .
$$
In the same way, we may construct higher-order approximants, which, 
however, we will not write down explicitly.

Let us stress that the self-similar root approximants (23) to (25) are 
valid for the whole range of variable (19), related to the coupling
parameter (7); they are also valid for an arbitrary anisotropy parameter
(4), incorporated in the variable (19), as well as for all quantum
numbers that enter expressions (23) to (25) through the parameters
(17) and the variable (19). The accuracy of the root approximants can
be  checked by comparing their values with those obtained numerically
from Eqs. (18) and (20).

Bose-Einstein condensates are often treated by using the Thomas-Fermi
approximation, when one omits the kinetic energy operator in Eq. (10).
In such a case, the solution to Eq. (10) reads
$$
\psi_{TF}(r,z) =\Theta\left ( r_0^2 - r^2 -\nu^2 z^2\right )\left (
\frac{r_0^2 - r^2 -\nu^2 z^2}{2g}\right )^{1/2} \; ,
$$
where $r_0^2\equiv 2E_{TF}$ and the energy is obtained from the
normalization condition for $\psi_{TF}$, which results in the
expression
$$
E_{TF} =\frac{1}{2}\left ( \frac{15}{4\pi} \; g\nu\right )^{2/5} \; .
$$
The Thomas-Fermi approximation is not self-consistent since the average
of the Hamiltonian (9), calculated with the wave function $\psi_{TF}$,
diverges. The energy $E_{TF}$ becomes correct for an asymptotically
large coupling parameter $g\ra\infty$, but for small and intermediate
$g$ this formula is not accurate. There have been several suggestions
of improving the Thomas-Fermi approximation by means of additional
corrections [33--35]. However, even after being improved, this 
approximation provides us solely the ground-state energy, being enable 
to describe the excited energy levels of the nonlinear coherent modes.

Figure 1 shows the ground-state energy of the condensate in a cigar-shape
trap, with $\nu=0.1$, calculated in different approximations. For
comparison, the Thomas-Fermi energy is also presented although its
accuracy up to the coupling parameters $g\approx 300$ is quite bad.
The maximal errors for the self-similar root approximants from
$E_1^*$ to $E_5^*$ are, respectively: $8\%,\; 3.5\%,\; 2\%,\; 1.2\%$, 
and $0.8\%$, which demonstrates good convergence.

The accuracy of the self-similar root approximants for different energy 
levels of the coherent modes, varying the coupling parameter $g$, and
for different trap shapes is illustrated in Figures 2 to 4, where the 
percentage errors as functions of $g$ are drown for a cigar-shape trap
(Fig. 2), spherical trap (Fig.3), and a disk-shape trap (Fig.4).
In all cases, there is the following standard situation. The maximal,
with respect to $g$ and quantum numbers, percentage errors for $E_1^*$ 
are between $4-12\%$; for $E_2^*$, between $2-5\%$, and for $E_3^*$, of
order $1\%$. The approximants $E_4^*$ and $E_5^*$ are close to $E_3^*$, 
because of which, not to overload the Figures, they are often omitted.

\section{Conclusion}

The nonlinear coherent modes of Bose-Einstein condensate at zero 
temperature are studied. These are described by the Gross-Pitaevskii 
equation and correspond to the stationary solutions of the latter. 
In the presence of a trapping potential, the spectrum of the coherent
modes is discrete. Our main aim in this paper was to find a convenient
and accurate way of calculating the energy levels of this spectrum for
arbitrary quantum numbers, any values of coupling parameters, and for
different aspect ratios of a cylindrical trap. One way for calculating 
the spectrum is by employing optimized perturbation theory, which 
requires numerical solution of the equations for control functions. 
Another possibility is to invoke the technique of self-similar root 
approximants, which results in sufficiently simple, and at the same 
time accurate, analytical formulas for the spectrum. The advantage 
of possessing analytical expressions is in the feasibility of a 
relatively easy study of their behaviour with respect to all parameters
characterizing the system. And to understand better the properties of 
the coherent modes is necessary for choosing the optimal conditions 
for their experimental observation and practical usage.

\vskip 5mm

{\bf Acknowledgement}

\vskip 2mm

The work was accomplished in the Research Center for Optics and 
Photonics, University of S\~ao Paulo, S\~ao Carlos. Financial support 
from the S\~ao Paulo State Research Foundation (FAPESP) is appreciated.

\newpage

\begin{center}
{\bf Figure Captions}
\end{center}

{\bf Fig. 1}. The ground-state energy for a cigar-shape trap, with 
$\nu=0.1$, as a function of the coupling parameter $g$, calculated by 
using formula (20) (solid line), and the self-similar root approximants 
$E_1^*$ (long-dashed line), $E_2^*$ (short-dashed line), and $E_3^*$
(dotted line). The energy $E_{TF}$ in the Thomas-Fermi approximation is
shown by the dashed-dotted line.

\vskip 1cm

{\bf Fig. 2}. The percentage errors of the self-similar root approximants 
for the energy levels of coherent modes in a cigar-shape trap, with
$\nu=0.1$, corresponding to $E_1^*$ (solid line),  $E_2^*$ (long-dashed 
line), $E_3^*$ (short-dashed line), $E_4^*$ (dotted line), and $E_5^*$ 
(dashed-dotted line) for different quantum numbers: (a) $n=m=j=0$; 
(b) $n=j=0,\; m=2$; (c) $n=j=0,\; m=10$.

\vskip 1cm

{\bf Fig. 3}. Percentage errors of the self-similar root approximants 
for $E_1^*$ (solid line),  $E_2^*$ (long-dashed line), and $E_3^*$ 
(short-dashed line) for a spherical trap, with $\nu=1$, and for 
different energy levels: (a) $n=m=j=0$; (b) $n=j=0,\; m=2$; 
(c) $n=3,\; m=2,\; j=1$.

\vskip 1cm

{\bf Fig. 4}. The case of a disk-shape trap with $\nu=10$. Percentage 
errors of the approximants $E_1^*$ (solid line),  $E_2^*$ (long-dashed 
line), $E_3^*$ (short-dashed line) and $E_4^*$ (dotted line) for different 
coherent modes: (a) $n=m=j=0$; (b) $n=j=0,\; m=2$; (c) $n=3,\; m=2,\; j=1$.


\newpage

\begin{thebibliography}{99}

\bibitem{1}
Parkins, A.S. and Walls, D.F., 1998, {\it Phys. Rep.}, {\bf 303}, 1.

\bibitem{2}
Dalfovo, F., Giorgini, S., Pitaevskii, L.P., and Stringari, S., 1999, 
{\it Rev. Mod. Phys.}, {\bf 71}, 463.

\bibitem{3}
Courteille, P.W., Bagnato, V.S., and Yukalov, V.I., 2001, {\it Laser Phys.}, 
{\bf 11}, 659.

\bibitem{4}
Yukalov, V.I., Yukalova, E.P., and Bagnato, V.S., 1997, {\it Phys. Rev. A},
{\bf 56}, 4845.

\bibitem{5}
Yukalov, V.I., Yukalova, E.P., and Bagnato, V.S., 2000, {\it Laser Phys.},
{\bf 10}, 26.

\bibitem{6}
Yukalov, V.I., Yukalova, E.P., and Bagnato, V.S., 2001, {\it Laser Phys.},
{\bf 11}, 455.

\bibitem{7}
Ostrovskaya, E.A. {\it et al}., 2000, {\it Phys. Rev. A}, {\bf 61}, 031601.

\bibitem{8}
Feder, D.L. {\it et al.}, {\it Phys. Rev. A}, {\bf 62}, 053606.

\bibitem{9}
Kivshar, Y.S., Alexander, T.J., and Turitsyn, S.K., 2001, {\it Phys. Lett. A},
{\bf 278}, 225.

\bibitem{10}
Williams, J.E. {\it et al.}, 1999, {\it Phys. Rev. A}, {\bf 61}, 33612.

\bibitem{11}
Woods, A.D.B. and Cowley, R.A., 1973, {\it Rep. Prog. Phys.}, {\bf 36},
1135.

\bibitem{12}
Steinhauer, J., Ozeri, R., Katz, N., and Davidson, N., 2001, {\it Preprint 
cond-mat/0111438}.

\bibitem{13}
Nepomnyashchy, Y.A., 1992, {\it Phys. Rev. B}, {\bf 46}, 6611.

\bibitem{14}
Svensson, E.C., Montfrooij, W., and de Schepper, I.M., 1996, {\it Phys. Rev. 
Lett.}, {\bf 77}, 4398.

\bibitem{15}
Yukalov, V.I., 1976, {\it Mosc. Univ. Phys. Bull.}, {\bf 31}, 10.

\bibitem{16}
Yukalov, V.I., 1976, {\it Theor. Math. Phys.}, {\bf 28}, 652.

\bibitem{17}
Yukalov, V.I., 1977, {\it Physica A}, {\bf 89}, 363.

\bibitem{18}
Yukalov, V.I., 1979, {\it Ann. Physik}, {\bf 36}, 31.

\bibitem{19}
Yukalov, V.I., 1980, {\it Ann. Physik}, {\bf 37}, 171.

\bibitem{20}
Caswell, W.E., 1979, {\it Ann. Phys.}, {\bf 123}, 153.

\bibitem{21}
Halliday, I. and Suranyi, P., 1980, {\it Phys. Rev. D}, {\bf 21}, 1529.

\bibitem{22}
Stevenson, P.M., 1981, {\it Phys. Rev. D}, {\bf 23}, 2916.

\bibitem{23}
Killingbeck, J., 1981, {\it J. Phys. A}, {\bf 14}, 1005.

\bibitem{24}
Fernandez, F.M. and Castro, E.A., 1982, {\it Phys. Lett. A}, {\bf 88}, 4.

\bibitem{25}
Feranchuk, I.D. and Komarov, L.I., 1982, {\it Phys. Lett. A}, {\bf 88}, 211.

\bibitem{26}
Okopi\'nska, A., 1987, {\it Phys. Rev. D}, {\bf 35}, 1835.

\bibitem{27}
Dineykhan, M., Efimov, G.V., Gandbold, G., and Nedelko, S.N., 1995, {\it
Oscillator Representation in Quantum Physics} (Berlin: Springer).

\bibitem{28}
Sissakian, A.N. and Solovtsov, I.L., 1999, {\it Phys. Part. Nucl.}, {\bf 30},
1057.

\bibitem{29}
Yukalov, V.I. and Yukalova, E.P., 1999, {\it Ann. Phys.}, {\bf 277}, 219.

\bibitem{30}
Marzlin, K.P. and Zhang, W., 1998, {\it Phys. Lett. A}, {\bf 248}, 290.

\bibitem{31}
Yukalov, V.I., Yukalova, E.P., and Gluzman, S., 1998, {\it Phys. Rev. A},
{\bf 58}, 96.

\bibitem{32}
Gluzman, S. and Yukalov, V.I., 1998, {\it Phys. Rev. E}, {\bf 58}, 4197.

\bibitem{33}
Dalfovo, F., Pitaevskii, L., and Stringari, S., 1996, {\it Phys. Rev. A},
{\bf 54}, 4213.

\bibitem{34}
Lundth, E., Pethick, C.J., and Smith, H., 1997, {\it Phys. Rev. A}, {\bf 55},
2126.

\bibitem{35}
Fetter, A.L. and Feder, D.L., 1998, {\it Phys. Rev. A}, {\bf 58}, 3185.

\end{thebibliography}
\end{document}